\documentclass[a4paper,11pt]{article}
\usepackage{fullpage,graphicx,xcolor}

\begin{document}

\title{Tag arrays}
\author{Travis Gagie}
\date{\today}
\maketitle

\begin{abstract}
The Burrows-Wheeler Transform (BWT) moves characters with similar contexts in a text together, where a character's context consists of the characters immediately following it.  We say that a property has contextual locality if characters with similar contexts tend to have the same or similar values (``tags'') of that property.  We argue that if we consider a repetitive text and such a property and the tags in their characters' BWT order, then the resulting string --- the text and property's {\em tag array} --- will be run-length compressible either directly or after some minor manipulation.
\end{abstract}

Mantaci, Restivo and Sciortino~\cite{MRS03} observed that the Burrows-Wheeler Transform (BWT) of a periodic text has most as many runs as the length of the period of the original text.  This is because the BWT turns {\em contextual} locality into {\em textual} locality, considering a character's context to be the characters that follow it in the original string.  For example, all of the {\tt G}s in the text in Figure~\ref{fig:periodic} are followed by {\tt ATTACAT\$} and are thus consecutive in the BWT.

\begin{figure}[b]
\begin{center}
{\tt \begin{tabular}{c@{\hspace{10ex}}c}
\begin{tabular}{c}
\textcolor{red}{GATTACAT\$} \\
\textcolor{blue}{GATTACAT\$} \\
\textcolor{green}{GATTACAT\$} \\
\textcolor{orange}{GATTACAT\$} \\
\textcolor{purple}{GATTACAT\$}
\end{tabular}
&
\begin{tabular}{c}
\textcolor{purple}{T}\textcolor{orange}{T}\textcolor{green}{T}\textcolor{blue}{T}\textcolor{red}{T} \\
\textcolor{purple}{T}\textcolor{orange}{T}\textcolor{green}{T}\textcolor{blue}{T}\textcolor{red}{T} \\
\textcolor{purple}{C}\textcolor{orange}{C}\textcolor{green}{C}\textcolor{blue}{C}\textcolor{red}{C} \\
\textcolor{purple}{G}\textcolor{orange}{G}\textcolor{green}{G}\textcolor{blue}{G}\textcolor{red}{G} \\
\textcolor{purple}{A}\textcolor{orange}{A}\textcolor{green}{A}\textcolor{blue}{A}\textcolor{red}{A} \\
\textcolor{purple}{\$}\textcolor{orange}{\$}\textcolor{green}{\$}\textcolor{blue}{\$}\textcolor{red}{\$} \\
\textcolor{purple}{A}\textcolor{orange}{A}\textcolor{green}{A}\textcolor{blue}{A}\textcolor{red}{A} \\
\textcolor{purple}{T}\textcolor{orange}{T}\textcolor{green}{T}\textcolor{blue}{T}\textcolor{red}{T} \\
\textcolor{purple}{A}\textcolor{orange}{A}\textcolor{green}{A}\textcolor{blue}{A}\textcolor{red}{A}
\end{tabular}
\end{tabular}}
\caption{A periodic string {\bf (left)} and its BWT {\bf (right)}, both written as matrices.}
\label{fig:periodic}
\end{center}
\end{figure}

The reverse --- that the BWT turns {\em textual} locality into {\em contextual} locality --- is not true if we consider as context succeeding characters in the BWT: the {\tt G} and {\tt A} at the start of the last copy of the repeated substring in the text (in purple) in Figure~\ref{fig:periodic} are adjacent but are followed by another {\tt G} and {\tt A} in the BWT, respectively.  It is true, however, if we consider as context the copies of the repeated substring from which those succeeding characters come (indicated by the colours in Figure~\ref{fig:periodic}): that {\tt G} and that {\tt A} from the last copy are both followed by an orange character, then a green character, then a blue character and then a red character.

Another way to say the BWT turns contextual locality into textual locality is that properties that have a run-like structure in {\em column}-major order when we write a periodic text as a matrix, have a run-like structure in {\em row}-major order in its BWT written as a matrix.  The most obvious of these properties are the identities of the characters themselves, and their positions in their copies of the repeated substring: the BWT of a periodic text with a relatively small period is run-length compressible.  If we measure only to the ends of the copies of the repeated substrings then the length of the longest common prefix (LCP) a suffix of the text shares with any other suffix --- that is, the distance from the start of that suffix to the next end of the copy of the repeated substring --- is another such property.  A fourth such property is the rather obscure interleaved LCP~\cite{GHKKNPS17} (ILCP) for the copies of the repeated substring, which measures to the ends of the copies of the repeated substring the LCP of a suffix and any other suffix of the same copy of the repeated substring.  For truly periodic strings, of course, we could say ``any other suffix of the repeated substring''; the ILCP was defined for collections of similar but different documents, however, and then it is important to say ``any other suffix of the same document''.

Symmetrically, another way to say the BWT turns textual locality into contextual locality is that properties that have a run-like structure in {\em row}-major order in the text have a run-like structure in {\em column}-major order in the BWT.  The most obvious of these properties is the copies of the repeated substring from which the characters come --- which has not only a run-{\em like} structure but consists of one run for each copy --- but there are several others: the positions of the characters are an incrementing sequence in row-major order in the text; if we measure to the end of the text then the LCPs (the {\em permuted} LCP array~\cite{KMP09}) are a decrementing sequence in row-major order in the text.

Truly periodic strings are common in combinatorics on words but in most applications of that field's results, strings are only approximately periodic.  For example, human genomes are very similar but not identical.  It is of interest, therefore, to study how well the observations above generalize to concatenations of similar strings, such as the toy alignment in Figure~\ref{fig:alignment} (with copies of {\tt -} representing skipped characters) and its BWT in Figure~\ref{fig:BWT}.  Unfortunately, experiments show that the run-length compressibility in column-major order of the LCPs measuring to the ends of strings does not scale to real pangenomic datasets.

\begin{figure}[t!]
\hspace{-3ex}
\begin{tabular}{c@{\hspace{5ex}}c}
{\tt \begin{tabular}{c}
-GATTACAT-\$ \\
AGAT-ACAT-\$ \\
-GAT-ACAT-\$ \\
-GATTAGAT-\$ \\
-GATTAGATA\# \\
\end{tabular}}
&
\begin{tabular}{rrrrrrrrrrr}
- & 1 & 2 & 3 & 4 & 5 & 6 & 7 & 8 & - & 10 \\
0 & 1 & 2 & 3 & - & 5 & 6 & 7 & 8 & - & 10 \\
- & 1 & 2 & 3 & - & 5 & 6 & 7 & 8 & - & 10 \\
- & 1 & 2 & 3 & 4 & 5 & 6 & 7 & 8 & - & 10 \\
- & 1 & 2 & 3 & 4 & 5 & 6 & 7 & 8 & 9 & 10
\end{tabular}
\\[15ex]
\begin{tabular}{rrrrrrrrrrr}
- & 5 & 4 & 3 & 6 & 5 & 4 & 3 & 2 & - & 1 \\
1 & 8 & 7 & 6 & - & 5 & 4 & 3 & 2 & - & 1 \\
- & 8 & 7 & 6 & - & 5 & 4 & 3 & 2 & - & 1 \\
- & 8 & 7 & 6 & 5 & 4 & 3 & 3 & 2 & - & 1 \\
- & 8 & 7 & 6 & 5 & 5 & 4 & 3 & 2 & 1 & 0
\end{tabular}
&
\begin{tabular}{rrrrrrrrrrr}
- & 0 & 2 & 1 & 1 & 1 & 0 & 2 & 1 & - & 0 \\
1 & 0 & 2 & 1 & - & 1 & 0 & 2 & 1 & - & 0 \\
- & 0 & 2 & 1 & - & 1 & 0 & 2 & 1 & - & 0 \\
- & 3 & 2 & 1 & 1 & 1 & 3 & 2 & 1 & - & 0 \\
- & 3 & 2 & 1 & 2 & 1 & 3 & 2 & 2 & 1 & 0
\end{tabular}
\\[15ex]
\begin{tabular}{rrrrrrrrrrr}
- & 0 & 0 & 0 & 0 & 0 & 0 & 0 & 0 & - & 0 \\
1 & 1 & 1 & 1 & - & 1 & 1 & 1 & 1 & - & 1 \\
- & 2 & 2 & 2 & - & 2 & 2 & 2 & 2 & - & 2 \\
- & 3 & 3 & 3 & 3 & 3 & 3 & 3 & 3 & - & 3 \\
- & 4 & 4 & 4 & 4 & 4 & 4 & 4 & 4 & 4 & 4
\end{tabular}
&
\begin{tabular}{rrrrrrrrrrr}
- &  0 &  1 &  2 &  3 &  4 &  5 &  6 &  7 &  - &  8 \\
9 & 10 & 11 & 12 &  - & 13 & 14 & 15 & 16 &  - & 17 \\
- & 18 & 19 & 20 &  - & 21 & 22 & 23 & 24 &  - & 25 \\
- & 26 & 27 & 28 & 29 & 30 & 31 & 32 & 33 &  - & 34 \\
- & 35 & 36 & 37 & 38 & 39 & 40 & 41 & 42 & 43 & 44
\end{tabular}
\\[15ex]
\multicolumn{2}{c}
{\begin{tabular}{rrrrrrrrrrr}
- &  3 &  2 & 1 & 2 & 1 & 0 &  1 &  0 & - & 0 \\
5 &  4 &  3 & 6 & - & 5 & 4 &  3 &  2 & - & 1 \\
- & 11 & 10 & 9 & - & 8 & 7 &  6 &  5 & - & 4 \\
- &  5 &  4 & 3 & 2 & 1 & 0 & 11 & 10 & - & 9 \\
- &  8 &  7 & 6 & 5 & 4 & 3 &  2 &  1 & 0 & 0
\end{tabular}}
\\[10ex]
\end{tabular}
\caption{A toy alignment {\bf (first row, left)} and its characters' column numbers in the alignment {\bf (first row, right)}, its LCP values measuring to the ends of the strings {\bf (second row, left)}, its ILCP values {\bf (second row, right)}, its characters' row numbers in the alignment {\bf (third row, left)}, its characters' positions in the concatenation of the strings {\bf (third row, right)}, and the PLCP values for the concatenation of the strings {\bf (fourth row)}.  The first three grids of numbers have run-like structure in column-major order and the second three grids of numbers have run-like structure in row-major order.}
\label{fig:alignment}
\end{figure}

\begin{figure}[t!]
\hspace{-3ex}
\begin{tabular}{c@{\hspace{10ex}}c}
\begin{tabular}{r|crrrrrr|rr}
 0 & \tt A  & 10 & 0 & 0 & 4 & 44 &  0 &  44 &  0 \\
 1 & \tt T  & 10 & 1 & 0 & 0 &  8 &  0 & -32 &  0 \\
 2 & \tt T  & 10 & 1 & 0 & 1 & 17 &  1 &   9 &  1 \\ 
 3 & \tt T  & 10 & 1 & 0 & 2 & 25 &  4 &   8 &  3 \\
 4 & \tt T  & 10 & 1 & 0 & 3 & 34 &  9 &   9 &  5 \\
 5 & \tt T  &  9 & 1 & 1 & 4 & 43 &  0 &   9 & -9 \\
 6 & \tt T  &  5 & 5 & 1 & 0 &  4 &  1 & -39 &  1 \\
 7 & \tt T  &  5 & 5 & 1 & 1 & 13 &  5 &   9 &  4 \\
 8 & \tt T  &  5 & 5 & 1 & 2 & 21 &  8 &   8 &  3 \\
 9 & \tt T  &  5 & 4 & 1 & 3 & 30 &  1 &   9 & -7 \\
10 & \tt T  &  5 & 5 & 1 & 4 & 39 &  4 &   9 &  3 \\
11 & \tt \$ &  0 & 1 & 1 & 1 &  9 &  5 & -30 &  1 \\
12 & \tt C  &  7 & 3 & 2 & 0 &  6 &  1 &  -3 & -4 \\
13 & \tt C  &  7 & 3 & 2 & 1 & 15 &  3 &   9 &  2 \\
14 & \tt C  &  7 & 3 & 2 & 2 & 23 &  6 &   8 &  3 \\
15 & \tt G  &  7 & 3 & 2 & 3 & 32 & 11 &   9 &  5 \\
16 & \tt G  &  7 & 3 & 2 & 4 & 41 &  2 &   9 & -9 \\
17 & \tt G  &  2 & 7 & 2 & 1 & 11 &  3 & -30 &  1 \\
18 & \tt G  &  2 & 7 & 2 & 2 & 19 & 10 &   8 &  7 \\
19 & \tt G  &  2 & 4 & 2 & 0 &  1 &  2 & -18 & -8 \\
20 & \tt G  &  2 & 7 & 2 & 3 & 27 &  4 &  26 &  2 \\
21 & \tt G  &  2 & 7 & 2 & 4 & 36 &  7 &   9 &  3 \\
22 & \tt A  &  6 & 4 & 0 & 0 &  5 &  0 & -31 & -7
\end{tabular}
&
\begin{tabular}{r|crrrrrr|rr}
23 & \tt A  &  6 & 4 & 0 & 1 & 14 &  4 &   9 &  4 \\
24 & \tt A  &  6 & 4 & 0 & 2 & 22 &  7 &   8 &  3 \\
25 & \tt A  &  6 & 3 & 3 & 3 & 31 &  0 &   9 & -7 \\
26 & \tt A  &  6 & 4 & 3 & 4 & 40 &  3 &   9 &  3 \\
27 & \tt A  &  1 & 8 & 0 & 1 & 10 &  4 & -30 &  1 \\
28 & \tt \$ &  1 & 8 & 0 & 2 & 18 & 11 &   8 &  7 \\
29 & \tt \# &  1 & 5 & 0 & 0 &  0 &  3 & -18 & -8 \\
30 & \tt \$ &  1 & 8 & 3 & 3 & 26 &  5 &  26 &  2 \\
31 & \tt \$ &  1 & 8 & 3 & 4 & 35 &  8 &   9 &  3 \\
32 & \tt A  &  8 & 2 & 1 & 0 &  7 &  0 & -28 & -8 \\
33 & \tt A  &  8 & 2 & 1 & 1 & 16 &  2 &   9 &  2 \\
34 & \tt A  &  8 & 2 & 1 & 2 & 24 &  5 &   8 &  3 \\
35 & \tt A  &  8 & 2 & 1 & 3 & 33 & 10 &   9 &  5 \\
36 & \tt A  &  8 & 2 & 2 & 4 & 42 &  1 &   9 & -9 \\
37 & \tt T  &  4 & 6 & 1 & 0 &  3 &  2 & -39 &  1 \\
38 & \tt A  &  3 & 6 & 1 & 1 & 12 &  6 &   9 &  4 \\
39 & \tt A  &  3 & 6 & 1 & 2 & 20 &  9 &   8 &  3 \\
40 & \tt T  &  4 & 5 & 1 & 3 & 29 &  2 &   9 & -7 \\
41 & \tt T  &  4 & 5 & 2 & 4 & 38 &  5 &   9 &  3 \\
42 & \tt A  &  3 & 3 & 1 & 0 &  2 &  1 & -36 & -4 \\
43 & \tt A  &  3 & 6 & 1 & 3 & 28 &  3 &  26 &  2 \\
44 & \tt A  &  3 & 6 & 1 & 4 & 37 &  6 &   9 &  3 \\
\multicolumn{10}{c}{}
\end{tabular}
\end{tabular}
\caption{The information from Figure~\ref{fig:alignment} but in BWT order instead of text order.  For ease of presentation on one page, we have also rotated the information 90 degrees --- so what should be run-like structure in row-major order according to our arguments is run-like structure in column-major order here.  The first column between the lines is the characters in the alignment in BWT order --- that is, the BWT itself --- and the other are the characters' columns in the alignment, the LCP values measuring to the end of the string, the ILCP, the characters' rows in the alignment, the characters' positions in the concatenation of the strings (now permuted into the suffix array), and the PLCP values (now permuted into the LCP array).  The columns to the right of the second line are the differentially compressed suffix array and LCP array.  The first four columns between the lines are visibly run-length compressible --- although this does not scale for the third column --- while the fifth and the two columns to the right of the second line display some repetitive structure.}
\label{fig:BWT}
\end{figure}

The run-length compressed BWT has been studied extensively (see, e.g.,~\cite{MNSV10,GNP20,KK20}) but otherwise properties with a run-like structure in row-major order have received more attention.  This is not surprising when we consider that the BWT turns the list of the strings from which the characters come into the {\em document array}, the characters' positions in the strings' concatenation into the {\em suffix array}, and the PLCP array into the {\em LCP array} --- three familiar and widely-used data structures in stringology.

This study has culminated in the theory of {\em string attractors}~\cite{KP18}.  If the BWT of the concatenation of $d$ documents has $r$ runs then the document array has a string attractor of size $O (r + d)$ and thus a straight-line program (SLP) with $O (r + d)$ rules.  The differentially compressed SA and LCP have string attractors of size $O (r)$ and thus straight-line programs (SLPs) with $O (r \log n)$ rules.  We differentially compress the SA and LCP because, mapped back to the characters' positions in the text and the PLCP, that turns incrementing and decrementing sequences into runs.

As far as we know, however, only the document array has a natural and direct interpretation in bioinformatics applications.  For example, it seems more interesting to know which column of the alignment a character in the BWT comes from --- a column-major property --- than to know its exact position in the concatenation of the strings (the corresponding entry in the SA minus 1, modulo the length of the text).  We admit that the other properties with contextual locality that we have mentioned so far are probably not of practical interest, but we argue there are others that are.

Bal\'az et al.~\cite{BGGHNPS24} recently considered characters' columns in an alignment and the vertices at which they are processed in a pangenome graph as properties with contextual locality.  For a property with contextual locality in text order, they called a character's value of that property its {\em tag}, and they called the list of the characters' tags in BWT order the {\em tag array} for that text and property (a name suggested by Shunsuke Inenaga).  Much of their paper was spent obtaining a worst-case upper bound for reporting the distinct tags in the runs of the tag array overlapping the BWT intervals of maximal exact matches (MEMs) between a pattern and the indexed text.  There is now a faster version of Li's forward-backward algorithm (see Algorithm 4 in~\cite{Li24}), however, that finds long MEMs and their BWT intervals and is much simpler and more practical than Bal\'az et al.'s approach.

Depuydt et al.~\cite{DRVVGF??} use a tag array of pangenomes of several species for metagenomic classification.  A character's tag in this case is the species of the genome containing that character.  Since the pangenomes consist of genomes from hundreds to thousands of individuals, nearly all characters have longer shared contexts with corresponding characters in other genomes of the same species than with characters in other species' genomes.  This means the property of ``species'' has contextual locality and the tag array is run-length compressible.

If there are fewer genomes from each species the tag array becomes less run-length compressible until, if there is only one genome from each species, we are left with a document array.  Even in that case, however, we conjecture that it may sometimes be worth considering contextual locality.  Specifically, if we have a phylogenetic tree for the species, then we can tag characters with the left-to-right positions in the phylogenetic tree of the genomes containing them.  It seems likely that nearly all characters have longer shared contexts with corresponding characters in closely related species than with characters in more distant species.  If so then the tag array will not be run-length compressible but neighbouring tags will tend to be similar, and thus the tag array will be differentially compressible.  This may be competitive with grammar-compressing the document array.

\end{document}